\documentclass[12pt]{article}

\usepackage[dvips]{graphicx}

\begin{document}

\title{\bf
Cluster computing performances using virtual processors and mathematical software}

\date{January 2004}

\author{\bf Gianluca Argentini \\
\normalsize gianluca.argentini@riellogroup.com \\
\textit{New Technologies and Models}\\
Information \& Communication Technology Department\\
\textit{Riello Group}, 35044 Legnago (Verona), Italy}

\maketitle

\begin{abstract}
In this paper I describe some results on the use of virtual processors technology 
for parallelize some SPMD computational programs in a cluster environment. 
The tested technology is the INTEL Hyper Threading on real processors, and the programs 
are MATLAB 6.5 Release 13 scripts for floating points computation. By the use of this
technology, I tested that a cluster can run with benefit a number of concurrent 
processes double the amount of physical processors.
The conclusions of the work concern on the utility and limits of
the used approach. The main result is that using virtual processors is a good technique
for improving parallel programs not only for memory-based computations, but in the case 
of massive disk-storage operations too.
\end{abstract}

\section{Introduction}
The processors virtualization technology permits to split a real physical processor
into two virtual chips, so that the operating system, as MS Windows or Linux, of a 
computer can use the virtual processors as two real chips. Example of such technology
is Intel's Hyper Threading [1]. The hardware can so be considered as a symmetric 
multi-processor machine and the software can use it as a true parallel 
environment.

  In this work I show some results obtained with parallel computations using 
Matlab [2] programs on Intel technology. A previous paper [3] describes the same cases
for an older Matlab version and for a single dual processor machine.
The physical and logical characteristics of 
the used cluster are presented in the following tables:\\

\begin{tabular}{|ll|}
\hline
\textbf{Hardware}		& \\
\hline
Type				&	2 nodes HP Compaq ProLiant DL360 \\
Processors	&	2 Intel Xeon 3.20 GHz for each node\\
Ram					&	2 GB for each node\\
Network			& 1 Gb switch for nodes connection\\
Storage			&	4 SCSI disks 36.5 GB - Raid 5 for each node\\
\hline
\end{tabular} \\ \\

\begin{tabular}{|ll|}
\hline
\textbf{Software}   & \\
\hline
Operating System		& MS Windows Server 2000 partition,\\
										& SuSE Linux 8.1 partition \\
Matlab							& v. 6.5.0 release 13 \\
\hline
\end{tabular} \\ \\

  The Matlab programs used for these experiments was based on cycles of floating-point 
computations.

\section{The parallel Matlab environment}
The package Matlab has not a native support for parallel elaboration and
multithreading [4]. Yet, there are some extensions, as tools and libraries [5], for
the use of a parallel environment on multi-processors hardware. With the cluster 
I have used the method of splitting a given computation on multiple instances
of the runtime Matlab program. A single master instance starts the slave copies on nodes
and  assigns to each of them the same set of instructions on different sets of data. 
Hence in the cluster I have simulated a SPMD computation. 

  In this way the parallel environment is simple, because there is not need of external 
libraries or calls to interfaces, and flexible, because to a single slave copy it can be 
assigned a set of different instructions for realizing a MPMD computation.

  With this method the exchange of messages among independent processes is a problem. 
The only way to communicate from one Matlab copy to another is the use of shared files. 
In a second type of experiments I show that this method is not critical for the 
time execution if one uses fast mass-storage as SCSI or FiberChannel systems, and the
nodes are connected in a fast private Lan.

\subsection{The SPMD programs}
In the experiments I have defined a \textit{master} Matlab function which writes 
to a shared file system the .m scripts to be executed by \textit{slaves} Matlab copies.
These copies are launched in background mode for the parallel execution. The master 
program controls the end of the computations using a simple set of lock-files. 
The slaves finish their work, save on files the results and cancel the own lock-file. 
The master reads the sets of data from these files for other possible computations. 
Now I describe the principal code of the program.\\
  
  This is the declaration of the function \textit{masterf}. The \textit{lockarray} 
variable is an array for testing the presence of the lock-files during the slaves 
computation. The \textit{finalres} is an array for the collection of the partial
results from slaves. The string \textit{computing} is the mathematical expression
to use in the computation. The array \textit{nodes} contains the names of the cluster's 
machines and it's used for the remote startup of the Matlab engines.

\scriptsize
\textbf{\\
function [elapsedtime,totaltime,executiontime]=masterf(nproc,maxvalue,step,computing)\\
\%\\
\% MASTERF: master function for parallel background computation.\\
\%\\
\% sintax:\\
\% [elapsedtime,totaltime,executiontime]=masterf(nproc,maxvalue,step,computing)\\
\%\\ 
\% input parameters:\\
\%\\
\% nproc = number of processes;\\
\% maxvalue = sup-limitation of the data-array to process; the inf-limitation is 0;\\
\% step = difference from two consecutive numbers in the data-array;\\
\% computing = the string of the mathematical expression to compute;\\
\%\\
\% output parameters:\\
\%\\
\% elapsedtime = total elapsed time to complete the execution of the computation;\\
\% totaltime = sum of the single slaves CPU-time to complete the single computation;\\
\% executiontime = single slaves CPU-time to complete the assigned computation;\\
\\
ostype=computer;\\
tottime=0.;\\
lockarray=0:nproc-1;\\
numbervalues=maxvalue;\\
computingstring=[' ' computing];\\
finalres=[];
nodes=[node01 node02];}\\

\normalsize
After the assignment of the own value to variable \textit{workdir}, working directory 
of Matlab, a cycle writes on storage the slaves lock-files.

\scriptsize
\textbf{\\
for i=0:nproc-1\\
\indent filelock = strcat(workdir,'filelock',int2str(i));\\
\indent fid=fopen(filelock,'wr');\\
\indent fwrite(fid,'');\\
\indent fclose(fid);\\
end}\\

\normalsize
In the next fragment of program, the master sets the commands for the writing 
of an appropriate Matlab .m script for every slave process. Such script contains
the instruction for determining the CPU-time spent on calculus, the expression 
of the mathematical computation, the instruction to save on storage the
data computed and the CPU-time, finally the instruction to delete the lock-file.

\scriptsize
\textbf{\\
for i=0:nproc-1\\
\indent  if (i==0) middlestep=0; else middlestep=1; end\\
\indent  infdata=i*(numbervalues/nproc) + middlestep*step;\\
\indent  supdata=(i+1)*(numbervalues/nproc);\\
\indent  fileworker = strcat(workdir,'fileworker',int2str(i),'.m');\\
\indent	 commandworkertmp = ...\\
\indent \indent      strcat('x=',num2str(infdata),':',num2str(step),':',num2str(supdata),...\\
\indent \indent      '; t1=cputime; ',computingstring,...\\
\indent \indent      '; t2=cputime-t1; save out',int2str(i));\\
\indent  commandworker = ['cd ' workdir '; ' commandworkertmp ...\\
\indent \indent      ' y t2; ' 'delete filelock'int2str(i) '; exit;'];\\
\indent  fid = fopen(fileworker,'wt');\\
\indent  fwrite(fid,commandworker);\\
\indent	 fclose(fid);\\
end}\\

\normalsize
The following instructions are OS-dependent, and are necessary for the right
setting of the command for remote startup of Matlab engines on nodes:

\scriptsize
\textbf{\\
switch ostype\\
case 'PCWIN'\\
\indent   osstring = 'dos';\\
\indent   workdir=strcat(matlabroot,'$\backslash$work$\backslash$');\\
\indent   startcommand='rcmd';\\
case 'LNX86'\\
\indent   osstring = 'unix';\\
\indent   workdir=strcat(matlabroot,'/work/');\\
\indent   startcommand='rsh';\\
end}\\

\normalsize
After the instructions for determining the CPU-time and the elapsed-time
(\textit{tic}) spent by the master program, a cycle launches the same number of 
slaves Matlab runtimes on each node. In the case of Windows operating system, 
the \textit{startcommand} string is "rcmd", the OS command for the background 
running of an executable program on a remote machine, and the \textit{osstring}
string is "dos". In the case of Unix-like operating system, the string are "rsh" and
"unix" respectively. Each slave executes immediately the \textit{fileworker} script, 
as shown by the Matlab "-r" parameter. The basic remote command is integrated by the
name of the node, alterning the order of startup for a simple reason of load 
balancing.

\scriptsize
\textbf{\\
t1 = cputime;\\
tic;\\
for i=0:nproc-1\\
\indent   if (mod(i,2)==0), startcommand = [startcommand node02]; else ... \\
\indent	\indent  startcommand = [startcommand node01]; end;\\
\indent   fileworker = strcat('fileworker',int2str(i));\\
\indent   commandrun = [startcommand ' matlab -minimize -r ' fileworker];\\
\indent   eval(strcat([osstring,'(','''',commandrun,'''',');']));\\
end}\\

\normalsize
In the next fragment of code the master program executes a cycle for determining
the end of slaves computations. It controls if the \textit{lockarray} variable has
some process's rank non negative. In this case, it attempts to open the relative 
lock-file; if the file still exists, the master closes it, else the lockarray 
process position is set to -1. The \textit{pause} instruction can be useful for
avoiding an excessive frequency, hence an high cpu-time consuming, in the "while"
cycle. 

\scriptsize
\textbf{\\
lockarraytmp=find(lockarray \begin{math}>\end{math} -1);\\
while (length(lockarraytmp) \begin{math}>\end{math} 0)\\
\indent   pause(.1);\\
\indent   for i=lockarraytmp\\		
\indent \indent      fid = fopen(strcat('filelock',int2str(i-1)),'r');\\
\indent \indent      if (fid \begin{math}<\end{math} 0)\\ 
\indent \indent \indent         lockarray(i) = -1;\\
\indent \indent      else\\
\indent \indent \indent       	fclose(fid);\\
\indent \indent      end\\
\indent  end\\
\indent  lockarraytmp=find(lockarray \begin{math}>\end{math} -1);\\
end}\\

\normalsize
At the end, the master reads the partial slaves computation outputs and stores 
them in an array.
At this point the master cpu-time and elapsed time are registered too. The total 
execution time is defined as sum of the slaves computation cpu-time, and is useful for
comparison with the execution time in the case \begin{math}\textit{nproc} = 1\end{math}.
The single slave execution time is defined as the arithmetic mean of all the partial
execution times.

\scriptsize
\textbf{\\
for i=0:nproc-1\\
\indent   partialres = load(strcat('out',int2str(i)));\\
\indent   finalres = [finalres partialres];\\
end\\
\\
elapsedtime = toc;\\
totaltime = cputime - t1;\\
\\
for i=0:nproc-1\\
\indent   tottime = tottime + partialres(i).t2;\\
\indent   executiontime = tottime/nproc;\\
end}\\

\normalsize

\section{Tests and results}
For the tests I have used the following values for the \textit{masterf} parameters:\\
\\
\textit{nproc}: from 2 to 16, step=2 (even numbers only, for right balancing\\
\indent of the nodes load);\\
\textit{maxvalue}: \textit{m} * 10000, where \textit{m} = 2, 4, 6;\\
\textit{step}: 0.001;\\
\textit{computing}: \begin{math} y = 5432.060708*\cos((\sin(x^{9.876}))^{-1.2345}) \end{math}.\\

I have also tested the program without the slaves saving of partial computations results
and their final master load, for determining the influence of the I/O storage operations
on the times of execution.\\

In the following table, the values are expressed in seconds. The number 2,...,16 are the
values of the \textit{nproc} parameter. I have not reported the 
elapsed-times, because they weren't different from the cpu-times registered, 
probably due to the fact that, during the experiments, the cluster was dedicated only 
to the computations.\\

In the case of no storage writing and reading of data results, the times are 20\%-30\% lower. 
The time values are those of MS-Windows case; in the Linux case the registered times are 
in general 15\%-20\% higher. This fact is probably due to a non optimized installation of
Linux distribution on nodes. \\ \\

\textbf{Table 1.} Total execution cpu-times, with data storage:\\

\begin{tabular}{|l|llllllll|}
\hline
\textbf{m} & \textbf{2} & \textbf{4} & \textbf{6} & \textbf{8} & \textbf{10}
& \textbf{12} & \textbf{14} & \textbf{16}\\
\hline
\scriptsize{2} & \scriptsize{48.29} & \scriptsize{27.70} & \scriptsize{32.51} & 
\scriptsize{22.56} & \scriptsize{28.14} & \scriptsize{31.34} & \scriptsize{33.28} &
\scriptsize{35.04}\\
\hline
\scriptsize{4} & \scriptsize{126.53} & \scriptsize{65.21} & \scriptsize{74.79} & 
\scriptsize{54.27} & \scriptsize{63.17} & \scriptsize{74.29} & \scriptsize{83.01} &
\scriptsize{91.34}\\
\hline
\scriptsize{6} & \scriptsize{263.37} & \scriptsize{109.48} & \scriptsize{121.30} & 
\scriptsize{78.41} & \scriptsize{116.23} & \scriptsize{125.69} & \scriptsize{138.51} &
\scriptsize{145.93}\\
\hline
\end{tabular} \\ \\

\section{Analysis of results}
From the results of the previous section, I deduce the following observations:\\
\begin{enumerate}
\item The case \textit{nproc}=8, hence the number of possible Hyper-Threading virtual 
processors based on the four physical chips, has the better performances for all the
values of the \textit{m} parameter;
\item The case \textit{nproc}=4, the number of physical processors in the cluster, has
a local peak of performances for all the values of the \textit{m} parameter;
\item The speedup [6] seems to be better for increasing values of the parameter \textit{m},
hence for larger amount of data to be computed; in the case \textit{m}=2 the speedup of
8 running processes over the case of 2 processes is about 2.14, while in the case
\textit{m}=6 the same speedup is about 3.35 (quasi-linear speedup).\\
\end{enumerate}

In the Fig. 1 the graphs are interpolations of the Table 1. data. The peaks of
performances at \textit{nproc}=4 and \textit{nproc}=8 are well visible, specially 
in the case \textit{m}=6.\\

\begin{figure}
\begin{center}
\includegraphics[width=10cm]{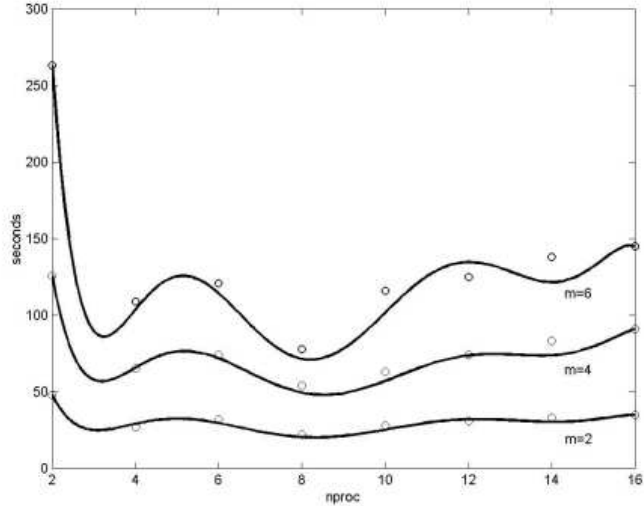}
\caption{Graphs of Table 1. data}
\end{center}
\end{figure}

\subsection{Conclusions}
From the previous facts one can deduce that a virtual processors technology 
as Hyper Threading on a cluster environment can be a good choice for running 
SPMD programs in the case that
\begin{itemize}
\item the number of parallel processes is equal to the number of virtual processors;
\item the data to be computed have a large amount, particularly when their distribution
among processes and the merging of final results are based on files stored on fast storage
system.
\end{itemize}

\section{Acknowledgements}
I wish to thank Simone Rossato, member of Infrastructures and Systems Area at 
ICT Department, Riello Group, for helpful discussions about the SuSE distribution of Linux,
and Dr. Marco Cavallone of MathWorks-Italy for the possibility of a free evaluation account
for Linux version of Matlab 6.5 .


\begin{thebibliography}{9}
\bibitem{1} \emph{www.intel.com/technology/hyperthread}, 2003\\
Web site for technical informations about Intel Hyper Threading technology.
\bibitem{2}	\emph{www.mathworks.com}, 2003\\
Web site of Mathworks, the producer of the mathematical package Matlab.
\bibitem{3} Gianluca Argentini, \emph{Using virtual processors for SPMD parallel programs}, 
www.arxiv.org/abs/cs.DC/0312049, 2003\\
The version of this work in the case of a single node and Matlab 5.3 .
\bibitem{4} \emph{www.mathworks.com/company/newsletter/pdf/spr95cleve.pdf}, 2003\\
This is a short but clear paper by Cleve Moler, co-founder of Mathworks, where the author
discusses why there isn't a parallel version of Matlab; the article has date 1995, but in
its essential philosophy is still valid.
\bibitem{5} \emph{www.mathtools.net/MATLAB/Parallel/index.html}, 2003\\
A Web page for a list of parallel extensions, as libraries and tools, to Matlab.
\bibitem{6} Peter Pacheco, \emph{Parallel programming with MPI}, Morgan Kaufmann, 1997\\
One of the best books for an introduction to parallel programming and its technical aspects;
it focuses on the Message Passing Interface.
\end{thebibliography}
\end{document}